# The Inverse Problem for Simple Liquid Metals a Study Case on Liquid Aluminum at Melting Point

Moustafa Sayem El-Daher*

**Abstract:**

In an attempt to test the possibility of solving the inverse problem for liquid metals i.e. obtaining the effective pair potential from the experimental structure factor, we solve the modified Hypernetted-Chain Integral equation for liquid aluminum at melting temperature to obtain the effective pair potential starting from the experimental structure factor and compare it with the potential obtained from theoretical considerations. Then we use the potential obtained by solving the inverse problem in Monte Carlo simulation to test it, and the calculated structure factor of the liquid aluminum is compared with experiment. We show that the solution of the inverse problem in such cases gives reasonable quantitative results, and reproduces the general features of the pair potential and the results for the structure factor are not far from the experimental measurements.

*****Assistant professor, Higher Institute for Laser Research and Applications, Damascus University.**

## 1- Introduction:

The structure of liquids is not easy to study since it lacks long range order which exists in solids, and thus must be studied using some statistical treatment which approximate the Liquid characteristics. Liquid structure is well described in physics by Integral equations derived from statistical mechanics treatment. Some of these equation are Percus-Yevick theory PY, Born-Green approximation BG, and Hypernetted-Chain theory HNC[1,2]; we will use a form of the latter in the current work. These equations relate the structure factor of the liquid (or equivalently the radial distribution function g(r)) to the effective pair potential which describe the interaction between the atoms within the liquid. If we know this potential we can calculate the structure factor of the liquid. These equations proved very helpful in dealing with liquids, depending of course on the model potential used in the calculations. These equations gave relatively accurate description in previous work[ 2,3,4] and in more recent studies[5,6,7,8].

In order to obtain the structure factor S(k) of the liquid or the radial distribution function which is related to it via a Fourier transform we need a pair potential representing the interaction between the particles within the liquid, and then solve the integral equation for S(k) or g(r). The pair potential can be a simple model potential like the hard sphere model or the charged hard sphere model[3,9] for simple cases. In the general case we need a more complicated potential to account for the many body interaction within the liquid, especially if the liquid particles are charged, then we need also to account for the long range Coulomb interaction between charged particles and the screening effects [10,11] of the free electrons, as it is the case in liquid metals. The problem thus becomes very complicated and need to be treated as a quantum many body problem. To tackle the problem we can use the screening function to account for the electron gas correlation obtained from the treatment of quantum field theory, random phase approximation[12], or using the fit through the total energy obtained from Quantum Monte Carlo method [13], or local density approximation within density functional theory [14-17], which makes the problem of finding the pair potential very demanding. It is therefore very tempting to take the integral equation and use an experimental structure factor (or radial distribution function) and solve the inverse problem to obtain the pair potential. But does it work?.

There are some positive results reported on the inverse problem, with some of it dealing with liquid metals [5,6,7,18], In the current work we use the experimental structure factor of liquid aluminum at melting point ( T=940 $^{o}$K) measured by X-ray diffraction as reported by Waseda[9], solve for the effective pair potential, use the resulting pair potential to calculate a theoretical structure factor using Monte Carlo simulation, and finally compare the results with experiment in order to check their validity and accuracy.

## 2- Goal of this study:

In order to explore the possibility and the accuracy of solving the inverse problem for liquid metals, we solved the inverse problem for liquid aluminum at melting temperature to obtain the pair potential, tested it in Monte Carlo simulation, and compared the results with other theoretically derived potentials. Previous studies were carried out mostly on mono-valant metals and were not tested through simulation. This work was done at the Higher Institute for Laser Research and Applications over a period of the last three years.

## 3- Theoretical Methods

By a "simple liquid" we mean one for which the interaction between particles can be assumed to be a two body central potential, $\phi(|\vec{r}_1 - \vec{r}_2|)$. Then, from statistical mechanics we have [1,2]

$$\nabla_{r1} g(r_{12}) = -\frac{g(r_{12})}{k_B T} \nabla_{r1} \phi(r_{12}) - \frac{\rho}{k_B T} \int dr_3 g^{(3)}(r_1, r_2, r_3) \nabla_{r1} \phi(r_{13}) \qquad (1)$$

We can express the pair distribution function $g(r_{12})$ in the form

$$g(r_{12}) = \exp(-U(r_{12})/k_B T) \qquad (2)$$

where $U(r_{12})$ is the potential of mean force. Inserting equation (2) in (1) we obtain

$$-\nabla_{r1} U(r_{12}) = -\nabla_{r1} \phi(r_{12}) - \rho \int dr_3 \frac{g^{(3)}(r_1, r_2, r_3)}{g(r_{12})} \nabla_{r1} \phi(r_{13}). \qquad (3)$$

The left hand side of the last equation represents the total force exerted on particle '1' by particle '2' located at distance $r_{12}$ from particle '1'. In the right hand side the force is expressed in two parts: first the direct interaction of particles '1' and '2' and secondly the effect of a third particle located at $r_{13}$ weighted by $g^{(3)}(r_1, r_2, r_3)/g(r_{12})$, where $g^{(3)}(r_1, r_2, r_3)$ is the three-body correlation function which expresses the probability of finding a third particle '3' at distance $r_{13}$ from the particle (1) assuming the presence of particles '1' and '2'.

Equation (3) which is called the force equation provides us with a way to calculate g(r) theoretically if we know the pair potential. The presence of $g^{(3)}(r_1, r_2, r_3)$ makes the calculation more complicated and we need to do some approximations.

A primitive approximation is the "superposition approximation" which expresses the function $g^{(3)}(r_1, r_2, r_3)$ as a product of pair correlation functions[1]:

$$g^{(3)}(r_1, r_2, r_3) = g(r_{12}) g(r_{23}) g(r_{31}). \qquad (4)$$

This approximation is the basis for some theories approximating the structure of liquids.

All liquid theories use the superposition approximation and introduce a number of correlation functions to describe the liquid state. The total correlation function h(r) defined as $h(r) = g(r) - 1$, which is a well behaved function which approaches zero when $r$ approaches infinity, and direct correlation functions defined as $c(r) = h(r) - 1$.

The Hypernetted-Chain theory HNC describes liquids by the following equations:

$$\begin{aligned}
g(r) &= \exp[\beta U(r) + N(r)] \\
N(r) &= h(r) - c(r) \\
h(r) &= g(r) - 1 \\
c(r) &= h(r) - \rho \int h(|r - r'|) c(r') . dr' \\
S(k) &= 1 + \rho \int h(r') . e^{ik.r} dr
\end{aligned} \qquad (5)$$

Where $g(r)$ is the radial distribution function, $N(r)$ is called the nodal function (obtained from statistical mechanics), $c(r)$ is the direct correlation function which is related to the total correlation function $h(r)$ by the formula

$$h(r) = c(r) + \rho \int c(|r - r'|) h(r') dr' \qquad (6)$$

$S(k)$ is the structure factor which can be obtained experimentally from x-ray diffraction. and we have[2]:

$$\frac{U - \phi}{kT} = -\rho \int c(r - r') h(r') dr' \qquad (7)$$

$$\frac{\phi_{HNC}(r)}{kT} = \frac{U(r)}{kT} + h(r) - c(r) \qquad (8)$$

Thus if we have S(k) from X-ray or neutron data, we can obtain $h(r)$ and $c(r)$ directly. Taking the Fourier transform of (6), we obtain

$$\tilde{h}(k) = \tilde{c}(k) + \tilde{h}(k)\tilde{c}(k) \qquad (9)$$

and hence

$$\tilde{c}(k) = \frac{\tilde{h}(k)}{1 + \tilde{h}(k)} = \frac{S(k) - 1}{S(k)}. \qquad (10)$$

Thus, the direct correlation function in k-space is directly related to the structure factor S(k).

The Modified Hypernetted-Chain theory MHNC adds the bridge correction B(r) based on statistical mechanics diagrammatic treatment of liquids, and we get the g(r) in the MHNC theory in the form:

$$g(r) = \exp[\beta U(r) + N(r) + B(r)] \qquad (11)$$

equation (6) becomes

$$h - c = \frac{\phi - U}{k.T} + B(r) \qquad (12)$$

**4- Results and discussion**

In the current work we use the MHNC theory to calculate the pair potential starting from the structure factor of liquid aluminum. The procedure is similar to that applied by Dharma et al [19]. We started from a parameterized form of a model potential of the form:

$$U(k) = Z^2 V(k) - \chi(k) V_{ie}^2(k) \qquad (13)$$

where: $\qquad V(k) = \dfrac{4\pi}{k^2},$

and:
$$V_{ie}(r) = \begin{cases} A_o, & r < R_o, \\ Z/r, & r \geq R_o, \end{cases} \qquad (14)$$

$A_0$ and $R_0$ are the parameters of the model potential and whose values are given in table (1). we have used Ashcroft hard sphere model for the bridge correction of the MHNC B(r) [20,21] , and $\chi(k)$ is dielectric response function of Geldert and Taylor[22,23,24].

We solve for the structure factor S(k) as mentioned in the previous section and optimize the parameters $A_0$ and $R_0$ in order to minimize the difference between the calculated and experimental S(k) taken at melting point of aluminum.

$$f = [S(k_i)_{calc} - S(k_i)_{expt}]^2 \qquad (15)$$

The experimental structure factor for the liquid aluminum which we used corresponds to temp 943 °K and to $r_0$=3.122 a.u. as given by x-ray diffraction measurements of Waseda[6], the data is in the range [0-5$k_f$]. Larger points are considered too noisy to include in the calculations. Fast Fourier transform was carried out using r grid in units of $r_0$ and the k grid in units of $r_0^{-1}$, where $r_0$ is the ion-sphere radius $r_0 = r_s Z^{-1/3}$, with $r_s$ is the Sietz radius. The calculation of the bridge function B(r) is done using appropriate hard sphere packing parameter η[20].

The results of the derived pair potential are given in Table (1) and a comparison with the ab-initio potential using GT screening function suggested by DRT [25] is given in Fig. (1).

Table (1): the numerical value obtained form inverting the MHNC equations.

| $r/r_o$ | pair potential |
|---|---|
| 1.0 | 78.420 |
| 1.2 | 29.242 |
| 1.4 | 6.962 |
| 1.6 | 0.870 |
| 1.8 | 0.479 |
| 2.0 | 0.449 |
| 2.2 | 0.301 |
| 2.4 | 0.087 |
| 2.6 | -0.033 |
| 2.8 | -0.039 |
| 3.0 | 0.00002 |
| 3.2 | 0.027 |
| 3.4 | 0.026 |
| 3.6 | 0.011 |
| 3.8 | 0.0012 |
| 4.0 | 0.001 |
| 4.5 | 0.0046 |
| 5.0 | 0.0061 |
| 5.5 | 0.0049 |
| 6.0 | 0.0005 |

The data the table is in units of $k_BT=0.0029862$ a.u

Table(2): Model potential parameters.

| A | -1.52207 |
|---|---|
| $R_0$ | 2.19716 |

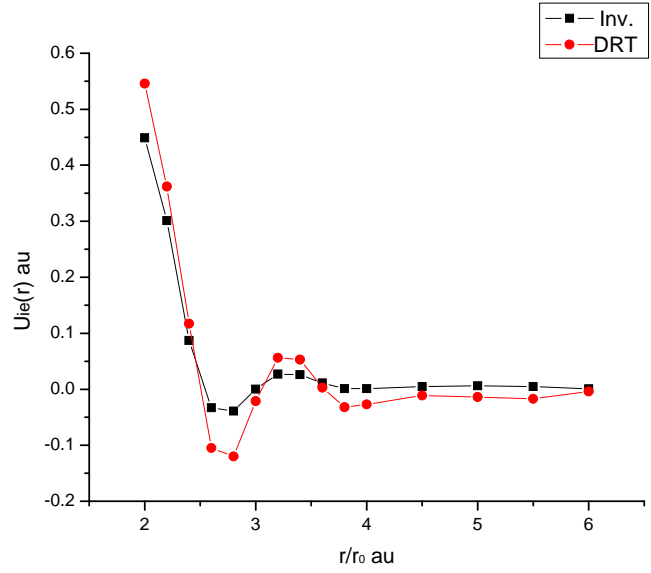

Fig (1): the derived potential compared with the theoretical DRT potential.

To check the accuracy of the pair potential obtained we use this potential in Monte Carlo simulation. We have used Metropolis algorithm [26,27] starting with an initial cubic cell with a side of 21 $\overset{o}{A}$ and a sample of 432 atoms distributed on (BCC) positions. The MC code used generates new configuration 300 times to eliminate any trace of the initial (BCC) distribution – this was checked in an earlier work and found satisfactory- before starting to accumulate statistical data in 2000 loops. The radial distribution function $g_{MC}(r)$ is obtained for $r \leq 21$ $\overset{o}{A}$. Moreover the asymptotic behavior of g(r) was taken into account by adding its contribution analytically to the calculation of the structure factor S(k) which is calculated from the equation:

$$S(k) = 1 + \rho \int [g(r) - 1] e^{i\bar{k}\bar{r}} dr \qquad (16)$$

The integration in the equation (16) is a three dimensional Fourier transform, we carried this integration out using the Filon method [28] for integration, The results of Monte Carlo simulation are given in Fig. (2) and compared with the experimental data of Waseda and to MC simulation obtained using ab intio potential based on Shaw's model and the VS screening function which was calculated using the same code is shown in Fig. (3).

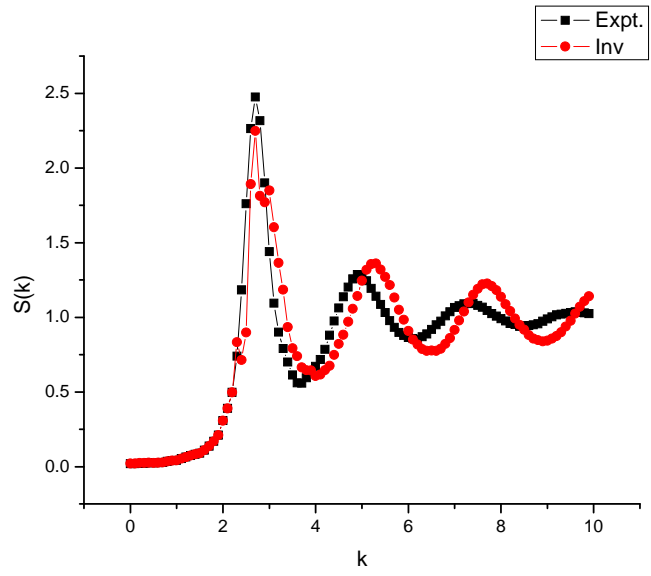

Fig (2): The structure factor S(k) of aluminum at melting point calculated by Monte Carlo method using the potential obtained in the current work.

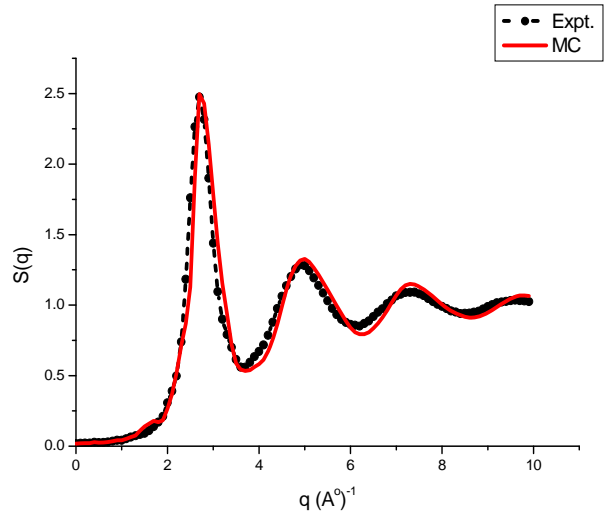

Fig (3): The structure factor S(q) of aluminum at melting point calculated by Monte Carlo method using DRT model potential.

Table (3): comparison of the peaks and positions of the obtained structure factor with the experimental and DRT theoretical structure factor.

|  | 1st peak | Position | 2nd peak | Position | Third peak | Position |
|---|---|---|---|---|---|---|
| Waseda Expt. | 2.475 | 2.7 | 1.265 | 5.1 | 1.093 | 7.4 |
| DRT | 2.496 | 2.7 | 1.301 | 5.1 | 1.1502 | 7.4 |
| Inverse problem | 2.248 | 2.7 | 1.361 | 5.3 | 1.225 | 7.7 |

We can see from Table(1) and Fig.(1) that although the pair potential derived from solving the inverse problem did reproduce the general features of the ab-initio potential of DRT but as we can see in Table(3) and Fig.(2) and Fig.(3) it gave less accurate structure factor when used in Monte Carlo simulations in comparison with the potential obtained from ab initio considerations.

5- **Conclusions:**

Despite carful attention to the accuracy of the numerical procedure used in the calculations and although the obtained potential from the inverse problem was reasonably similar to the potential obtained from theoretical considerations, the results of the Monte Carlo simulation were less accurate than other potentials. In our view this shortcoming comes primarily from the approximations inherent to the integral equations, and therefore a further refinement is needed.

A promising approach is the use of predictor-corrector scheme which correct for the predictors –the integral equation in this case- with a molecular dynamics simulation which is done for each step which makes the calculations very demanding [18,29,30]. This approach needs however further testing to explore a range of temperatures especially near critical points for different liquid metals.